\documentclass[
 aip,
 jmp,
 amsmath,amssymb,
 reprint,
]{revtex4-1}

\usepackage{graphicx}
\usepackage{dcolumn}
\usepackage{bm}
\usepackage[toc,page]{appendix}
\usepackage{xcolor}

\begin{document}

\setcounter{page}{1}

\title{Thermo-elasto-plastic simulations of femtosecond laser-induced structural modifications: application to cavity formation in fused silica}

\author{Romain Beuton}
 \email{romain.beuton@u-bordeaux.fr}
 \affiliation{Universit\'e de Bordeaux-CNRS-CEA, Centre Lasers Intenses et Applications, UMR 5107, 33405 Talence, France}
\author{Beno\^it Chimier}%
 \affiliation{Universit\'e de Bordeaux-CNRS-CEA, Centre Lasers Intenses et Applications, UMR 5107, 33405 Talence, France}
\author{J\'er\^ome Breil}
 \affiliation{CEA/CESTA, 15 Avenue des Sabli\`eres, CS 60001 33116 Le Barp cedex France}
\author{David H\'ebert}
 \affiliation{CEA/CESTA, 15 Avenue des Sabli\`eres, CS 60001 33116 Le Barp cedex France}
\author{Pierre-Henri Maire}
 \affiliation{CEA/CESTA, 15 Avenue des Sabli\`eres, CS 60001 33116 Le Barp cedex France}
\author{Guillaume Duchateau}%
 \affiliation{Universit\'e de Bordeaux-CNRS-CEA, Centre Lasers Intenses et Applications, UMR 5107, 33405 Talence, France}

\begin{abstract}
The absorbed laser energy of a femtosecond laser pulse in a transparent material induces a warm dense matter region which relaxation may lead to structural modifications in the surrounding cold matter.
The modeling of the thermo-elasto-plastic material response is addressed to predict such modifications. It has been developed in a 2D plane geometry and implemented
in a hydrodynamic lagrangian code. The particular case of a tightly focused laser beam in the bulk of fused silica is considered as a first application
of the proposed general model. It is shown that the warm dense matter relaxation, influenced by the elasto-plastic behavior of the surrounding cold matter, generates both a strong shock
and rarefaction waves. Permanent deformations appear in the surrounding solid matter if the induced stress becomes larger than the yield strength. This interaction results
in the formation of a sub-micrometric cavity surrounded by an overdense area. This approach also allows one to predict regions where cracks may form. The present modeling
can be used to design nano-structures induced by short laser pulses.
\end{abstract}

\maketitle 

\section{INTRODUCTION}
Femtosecond laser pulses are widely used to structure transparent dielectric materials by modifying their local properties, resulting in the formation of designed structures 
on the surface or in the bulk \cite{Davis1996, Shimotsuma2003, Gattass2008, Gottmann2009, Cheng2009, Bonse2012, Richter2012, Buividas2014, Liao2015}. 
This leads to various industrial and technological applications as optical data storage, waveguide or grating writing \cite{Courvoisier2013, Courvoisier2016, Bellouard2016}.
In the case of femtosecond laser pulses tightly focused in the bulk, relatively high intensities, above $10^{14}$ W/cm$^2$, can be reached. Such intensities enable to ionize most of the transparent dielectric materials
(even those exhibiting large optical bandgaps) by inducing nonlinear electronic responses. 
Because the laser beam deposits its energy on a timescale smaller than hydrodynamic characteristic times (of the order of tens picoseconds), there is a decorrelation between the energy absorption processes and 
the evolution of macroscopic matter properties.
A warm dense matter with both a high temperature and pressure is then created in the focal volume of characteristic dimensions of the order of a few tenths of micrometer. This area is surrounded by a non-modified cold material.
The relaxation of this heated matter generates a strong shock and rarefaction waves which propagate through the solid. 
For energy densities above the material damage threshold, a significant structural modification may appear by compression
and traction of the material. In particular, several studies and experimental observations have shown 
that a single ultrashort laser pulse could form a cavity in the bulk material \cite{Juodkazis2003, Juodkazis2006, Gamaly2008}.

Important efforts and significant progresses have been performed to understand the physical processes at play during the laser-material interaction including the solid phase response
\cite{Gamaly2006, Hallo2007, Sakakura2007, Mezel2008, Hebert2011, Bulgakova2015, Najafi2016}. Especially, the recent work by Najafi et al.\cite{Najafi2016} investigates the birefringence induced by the axisymmetric stress
due to the interaction of an ultrashort laser with a transparent material.
In the particular case of tightly focused beams
theoretical and numerical studies have been carried out by only accounting for the hydrodynamic behavior \cite{Gamaly2006, Hallo2007, Mezel2008} (fluid behavior) or by including the solid
mechanical properties \cite{Hebert2011, Bulgakova2015}. In the latter study, despite only small material density variations have been considered, it has been shown that the behavior of the 
solid is very important and play a significant role in the formation of the desired structures. Based on these considerations, it appears that theoretical and numerical developments are still desired to
model and understand significant laser induced material modifications by tightly focused beam.

The aim of this paper is to model these laser-induced modifications and understand the transient mechanisms leading to a cavity formation. The solid behavior is introduced in the hydrodynamic Euler's equations.
Generally, materials exhibit a viscoelastic behavior mixing both the elastic properties of solids and the viscosity one of liquids \cite{Christensen1971, Meyers2009}.
For the present applications, only small material deformations are assumed since experimental observations show that the overdense region around the cavity is significantly smaller than the size of
the total modified area \cite{Gamaly2008}. The material response thus mainly exhibits an elastic behavior where the viscosity influence may be neglected. However, due to these deformations, when
the induced mechanical stress exceeds the elastic limit, permanent deformations are produced. Within this framework, the solid material response then consists of
two regimes of deformation: an elastic (reversible) regime and a plastic (irreversible) regime.

In the present work, to describe the elasto-plastic (EP) response of the solid, the physical model proposed by Wilkins \cite{Wilkins1964} is used (Section \ref{th.modeling}).
It has been implemented in the hydrodynamic CHIC code \cite{Breil2010, Maire2013} 
to study the cavity formation by a tightly focused laser beam (Section \ref{results}).
The simulations are performed for fused silica by assuming an instantaneous energy deposition. Throughout the paper, the EP predictions are compared to fluid simulations. First, 1D simulations are considered in order to
exhibit the influence of the EP behavior. Especially, a difference between the shock waves (shape and velocity) clearly appears, leading to a significant discrepancy between the density profiles.
Results of 2D simulations demonstrate the role of the solid response for the stabilization of the laser-induced nano-structure, providing caracteristic deformations in a good agreement with experimental observations. 
Finally, a prediction of the critical zones 
where potential fractures may appear is proposed. Conclusions and perspectives are drawn in Section \ref{conclusion}.

\section{THEORETICAL MODELING}
\label{th.modeling}
Following references [\onlinecite{Wilkins1964}] and [\onlinecite{Maire2013}], the relaxation of the heated matter in the surrounding colder material can be described by a more general form of the hydrodynamic conservation laws, 
where the pressure $P$ is substituted by the Cauchy
stress tensor \cite{Irgens2008} $\bar{\bar\sigma}$. Within their lagrangian form, this set of equations reads: 
\begin{equation}
 \rho\frac{d}{dt}(\frac{1}{\rho})-\nabla.\textbf{V}=0
\end{equation}

\begin{equation}
 \rho\frac{d\textbf{V}}{dt}-\nabla.\bar{\bar\sigma}=\textbf{0}
\end{equation}

\begin{equation}
 \rho\frac{dE}{dt}-\nabla.(\bar{\bar\sigma}.\textbf{V})=0
\end{equation}
where $\rho$, $\textbf{V}$ and $E$ are the density, the velocity and the specific total energy, respectively.
The Cauchy stress tensor consists of two terms:
\begin{equation}
 \bar{\bar\sigma}=-PI_d+\bar{\bar S} 
 \label{sigma}
\end{equation}
The first term of \ref{sigma} corresponds to the spherical part of the stress tensor, representing the fluid behavior of matter, which only changes the volume of the material when a pressure
is applied. $I_d$ is the identity matrix. The second term, $\bar{\bar S}$, is the deviatoric part of 
the stress tensor accounting for the solid behavior. It includes both longitudinal and shear stresses.
The time evolution of the deviatoric stress is given by \cite{Maire2013}:
\begin{equation}
 \frac{d\bar{\bar S}}{dt} = 2\mu(\bar{\bar D}_0-\bar{\bar D}^p)-(\bar{\bar S}\bar{\bar W}-\bar{\bar W}\bar{\bar S})
\label{eq:dSdt}
\end{equation}
where $\mu$ is the shear modulus characterizing how the material deforms under the influence of a shear stress, and $\bar{\bar W}$ is the antisymmetric part of the velocity gradient
\begin{equation}
 \bar{\bar W}=\frac{1}{2}[\nabla.\textbf{V}-(\nabla.\textbf{V})^t]
\end{equation}
$\bar{\bar D}_0$ is the deviatoric part of the strain rate tensor $\bar{\bar D}$,
\begin{equation}
 \bar{\bar D}_0=\bar{\bar D}-\frac{1}{3}Tr(\bar{\bar D})I_d
\end{equation}
with $\bar{\bar D}$ defined as the symmetric part of the velocity gradient:
\begin{equation}
 \bar{\bar D}=\frac{1}{2}[\nabla.\textbf{V}+(\nabla.\textbf{V})^t]
\end{equation}
The plastic strain rate $\bar{\bar D}^p$ in (\ref{eq:dSdt}) is determined through the equation:
\begin{equation}
 \bar{\bar D}^p=\chi(\bar{\bar N}^p:\bar{\bar D}^p)\bar{\bar N}^p
\end{equation}
where the symbol : denotes the inner product of tensors defined as $\bar{\bar A}:\bar{\bar B}=Tr(\bar{\bar A}^t\bar{\bar B})$.  
$\bar{\bar D}^p$ represents the rate of deformation due to the formation and motion of dislocations in the material.
$\bar{\bar N}^p$ is the plastic flow direction
\begin{equation}
 \bar{\bar N}^p=\frac{\bar{\bar S}}{|\bar{\bar S}|}
\end{equation}
and $\chi$ is the switching parameter defined by
\begin{equation}
 \chi = \begin{cases} 
	      0 & \text{if } f<1 \text{ or if } f=1 \text{ and } (\bar{\bar N}^p:\bar{\bar D}^p)\leq0 \\ 
	      1 & \text{if } f=1 \text{ and } (\bar{\bar N}^p:\bar{\bar D}^p)>0       
	\end{cases}
\end{equation}
where $f$ is the yield function. The first case corresponds to different regimes except the plastic domain: the elastic regime ($f<1$), elastic unloading ($f=1$ and $(\bar{\bar N}^p:\bar{\bar D}^p)<0$) 
and neutral loading ($f=1$ and $(\bar{\bar N}^p:\bar{\bar D}^p)=0$). The second case corresponds to the plastic regime.

The evolution of the deviatoric stress $\bar{\bar S}$ is thus subjected to the yield function allowing to determine the deformation regime of the solid submitted to a mechanical stress.
In the present study the yield function is defined by the von Mises yield criterion \cite{Mises1913}, which is widely used.
In this case, the yield function reads:
\begin{equation}
 f=\frac{\sigma_{eq}}{Y} 
\end{equation}
 where $Y$ defined the yield strength of the material which represents its elastic limit, and  $\sigma_{eq}$ is a local effective stress, called the equivalent stress.
 \begin{equation}
 \sigma_{eq} =\sqrt{\frac{3}{2}Tr(\bar{\bar S}.\bar{\bar S})}
\end{equation}

Within the plastic regime, a 
phenomenon of hardening may take place increasing the flow resistance (the yield strength) \cite{Fanchon2001}. However, in this study, a perfect plastic behavior is considered corresponding to the ideal 
case where the yield strength remains constant. This assumption is correct for a brittle material \cite{Irgens2008} like fused silica which is studied in Section \ref{results}.

Within the elastic regime, the flow velocity can be decomposed into a longitudinal and a transverse components. 
Each direction may support waves which velocities $c_{l}$ and $c_{t}$, respectively, read \cite{Royer2000}:
\begin{equation}
\label{vit_l}
 c_{l}^2 = c_{h}^2 + \frac{4}{3}\frac{\mu}{\rho}
\end{equation}
and
\begin{equation}
 c_{t}^2 = \frac{\mu}{\rho}
\end{equation}
where $c_{h}$ is the standard hydrodynamic sound velocity. Note that the longitudinal sound velocity (Eq. (\ref{vit_l})) is larger in a solid than in a fluid due to the lattice structure.
Within the plastic regime, only compression waves can propagate since the solid structure is removed for large stresses, \textit{i.e.} $c_{l} = c_{h}$ and $c_{t} = 0$.

Finally, a thermal softening is introduced in the model to account for the solid-liquid phase transitions. This is a general model where a polynomial $g(\varepsilon)$, which is a function of the specific internal energy $\varepsilon$, 
weights the shear modulus and the yield strength.
It is equal to unity and zero for the solid and liquid phases, respectively. This polynomial accounts for a smooth transition between the two phases (see Appendix A for more details).

Based on the previously presented model, a second order cell-centered Lagrangian numerical scheme is implemented in the hydrodynamic CHIC code \cite{Breil2010}.
This scheme possesses several efficient numerical features \cite{Maire2013}. It is developed in a planar geometry
allowing one to study more accurately physical phenomena than in a 1D geometry as presented hereafter.

\section{RESULTS AND DISCUSSION}
\label{results}
The cavity generation in the bulk of fused silica is considered. Such a structure may be induced by tightly focusing a femtosecond laser pulse into the material. During the laser-matter interaction, nonlinear processes lead to a significant absorption of 
the laser energy.
The latter is then confined
in a volume with characteristic dimensions smaller than the laser wavelength. For absorbed energies of the order of 10 nJ, 
the induced high energy densities create high enough
temperatures and pressures, the latter exceeding the bulk modulus of the solid, to form a void with observed dimensions between 0.2 $\mu$m and 0.5 $\mu$m. 
In the present work, a gaussian energy deposition is assumed in order to simplify the study of the dynamics of cavity formation. 
The energy deposition radius \cite{Gamaly2006} is set to 0.13 $\mu$m at half maximum. This radius is chosen to remain constant even for higher energy of the laser beam where the volume of absorption may be larger in experiments. 
This work is motivated by the fact that the main aim of the present study is to demonstrate the importance
of the solid response. A more detailed study, including an initial energy deposition evaluated by the solving of the Maxwell's equations, will be provided elsewhere. The absorbed 
energy density is chosen (value provided hereafter) such that the simulation leads to cavity sizes similar to the experimental observations; 0.4 $\mu$m is chosen in particular. 
It is noteworthy that the regime leading to such cavity radii, as shown hereafter, exhibits 
the influence of most of the possible physical mechanisms for such a system.

An equation of state (EOS) is required to obtained the pressure and the temperature from the knowledge of the density and the specific internal energy. For that purpose, the SESAME table 7386 \cite{Boettger1989} for fused silica is used.
The initial density is equal to 2.2 g/cm$^3$.
The yield strength and the shear modulus for initial standard conditions are $Y_0=7.1$ GPa, $\mu_0=22.6$ GPa, respectively. 
Within the present hypothesis of an ideal plastic behavior, $Y_0$ remains constant whatever the deformation. This assumption is correct for SiO$_2$ since it is a brittle material \cite{Irgens2008}.
Despite fused silica can present non ideal plastic deformations at the micron scale \cite{Kermouche2008}, the plasticity domain remains relatively short.

The thermal conductivity $\lambda_{th}=1.381$ W.m$^{-1}$.K$^{-1}$ is also assumed to be constant. Indeed, within the nanosecond timescale ($t=1$ ns) corresponding to the simulation time, with
a specific heat capacity $C$ of 1000 J.kg$^{-1}$.K$^{-1}$, the characteristic length of diffusion $l$
is about 0.025 $\mu$m following the relation $l=\sqrt{\lambda t/\rho C}$. This is smaller than the characteristic lengths of deformations studied here, making the influence of the thermal conductivity negligible for the present study.
Moreover, we have checked that higher values of the thermal conductivity do not change the results on this timescale.

Note that the solid-solid phase transitions are also neglected.
As mentioned in [\onlinecite{Michel2006}], for stresses above 35 GPa, a change in slope of the Hugoniot curve is observed due to a 
crystallization toward the stichovite phase of SiO$_2$. However, the time required for the matter to undergo this phase transition is of the order of 1 to 10 ns depending on the strength of the shock \cite{Gleason2015}. 
For the present studies, this timescale is significantly longer than the time for which the matter experiences the required conditions when the shock passes through. 
Moreover, experimental observations have never revealed, to our knowledge, the stishovite phase in fused silica after an interaction with a femtosecond laser, further supporting the present assumption.

\subsection{Comparison of the fluid model and the elasto-plastic model in the 1D case}
\label{sec:hydro_vs_elasto}
Let us first considered a 1D simulation to compare the EP model to the fluid description in order to exhibit the influence of the solid response.
The initial absorbed energy density is set to 90 kJ/cm$^3$ (0.9 nJ) in the center of the 1D domain. The full cartesian mesh size is set to 20 $\mu$m with initial cell length set to 20 nm.   

\begin{figure}[!h]
  \centering
  \includegraphics[width=8cm]{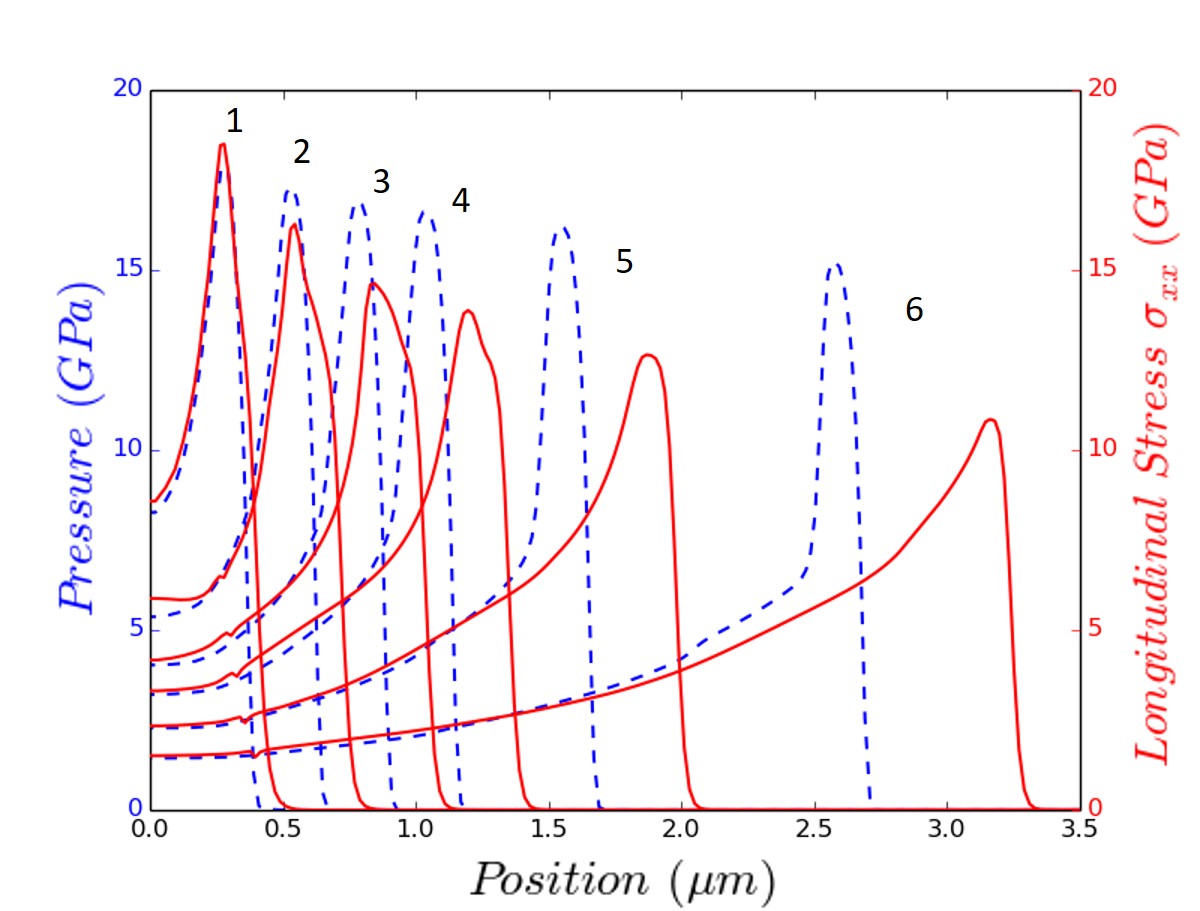}
  \caption{Spatial profiles of the pressure and the longitudinal stress $\sigma_{xx}$ as predicted by the fluid model (dashed lines) and the EP model (lines), respectively, at
  50, 100, 150, 200, 300, 500 ps (curves 1-6, respectively). Simulations are performed in the 1D case for an absorbed energy density of 90 kJ/cm$^3$.}
  \label{Figure_1}
\end{figure}
For various times, Fig. \ref{Figure_1} presents the spatial evolution of the pressure (fluid model) and the stress $\sigma_{xx}$ (EP model). We remind that the stress is a generalization of 
the pressure within the EP model, the comparison thus makes sense. The stress is defined as positive in compression for the sake of clarity. 

The propagation of a shock, induced by
the relaxation of the heated matter, is observed in both cases but with different shapes and velocities.
In the fluid model case, the shock propagates with a straight front without change in its shape as usually. Only its amplitude decreases in course of propagation.
In the case of the EP model, the shock propagates similarly as in the pure fluid model up to 100 ps. However, from 100 ps, a change is observed in the shape of the shock wave and disappears after 300 ps. 
This behavior is due to the surrounding matter which is in a solid state. The matter is submitted to strong stresses and the EP behavior split the shock into two
waves: a plastic wave and an elastic precursor. The first, inducing permanent deformations in the material, propagates with the hydrodynamic speed of sound $c_{h}$, itself depending on the local density value.
The second, inducing reversible deformations, propagates with the augmented longitudinal speed of sound $c_{l}$ (Eq. (\ref{vit_l})). Its amplitude is equal to the Hugoniot elastic limit (HEL) $Y_{HEL}$ 
which is around 12 GPa as provided by both the SESAME table and the mechanical properties of the material, following the relation \cite{Rosenberg1993}: $Y_{HEL}=\frac{1-\nu}{1-2\nu}Y_0$ with $\nu$ the Poisson's ratio (around 0.3-0.4 for glass). 
Note that the HEL is slightly different from the yield strength.
It defines the threshold in longitudinal stress where there is a transition between a pure elastic behavior to an elastic-plastic one \cite{Asay1993}.
The elastic wave then is faster than the plastic wave explaining the observed splitting. Finally, after 300 ps, only an acoustic wave propagates because $\sigma_{xx}<Y_{HEL}$ ($\sigma_{eq}<Y$), permanent deformations in compression then are not further allowed.

\begin{figure}[!h]
  \centering
  \includegraphics[width=8cm]{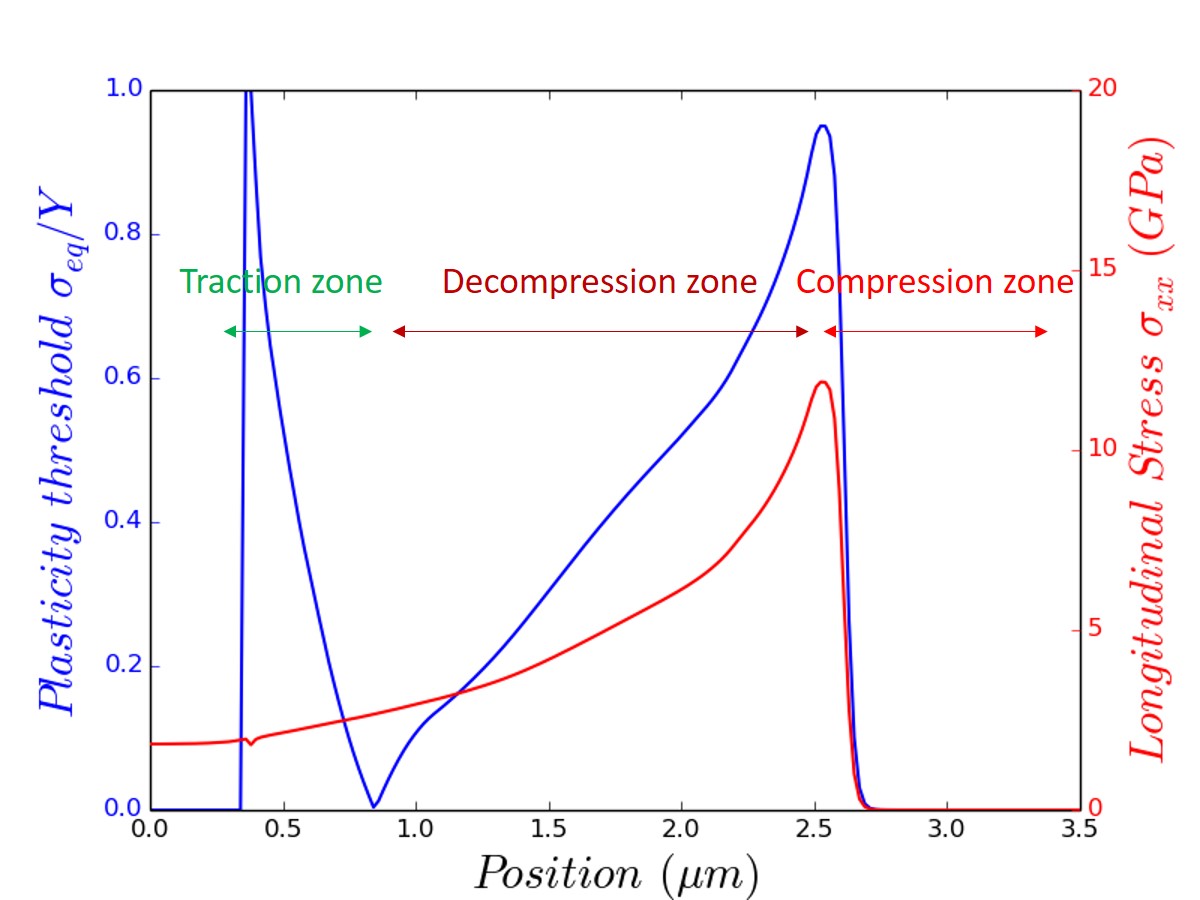}
  \caption{Structure of the shock wave and associated plasticity threshold curve at 400 ps.}
  \label{Figure_2}
\end{figure}
With this particular two-wave structure, the shock front first compresses the matter elastically and then plastically. The previously compressed material is then decompressed by the back of
the wave, and finally is stretched (sign change in the deviatoric stress) after the wave. 
If this traction
is strong enough, it may deform plastically the matter a second time. This scenario is illustrated by Fig. \ref{Figure_2} which shows the stress profile at 400 ps.
The second plastification is demonstrated by the plotted plasticity threshold curve that provides the value of $\sigma_{eq}/Y$ for each position.

\begin{figure}[!h]
  \centering
  \includegraphics[width=8cm]{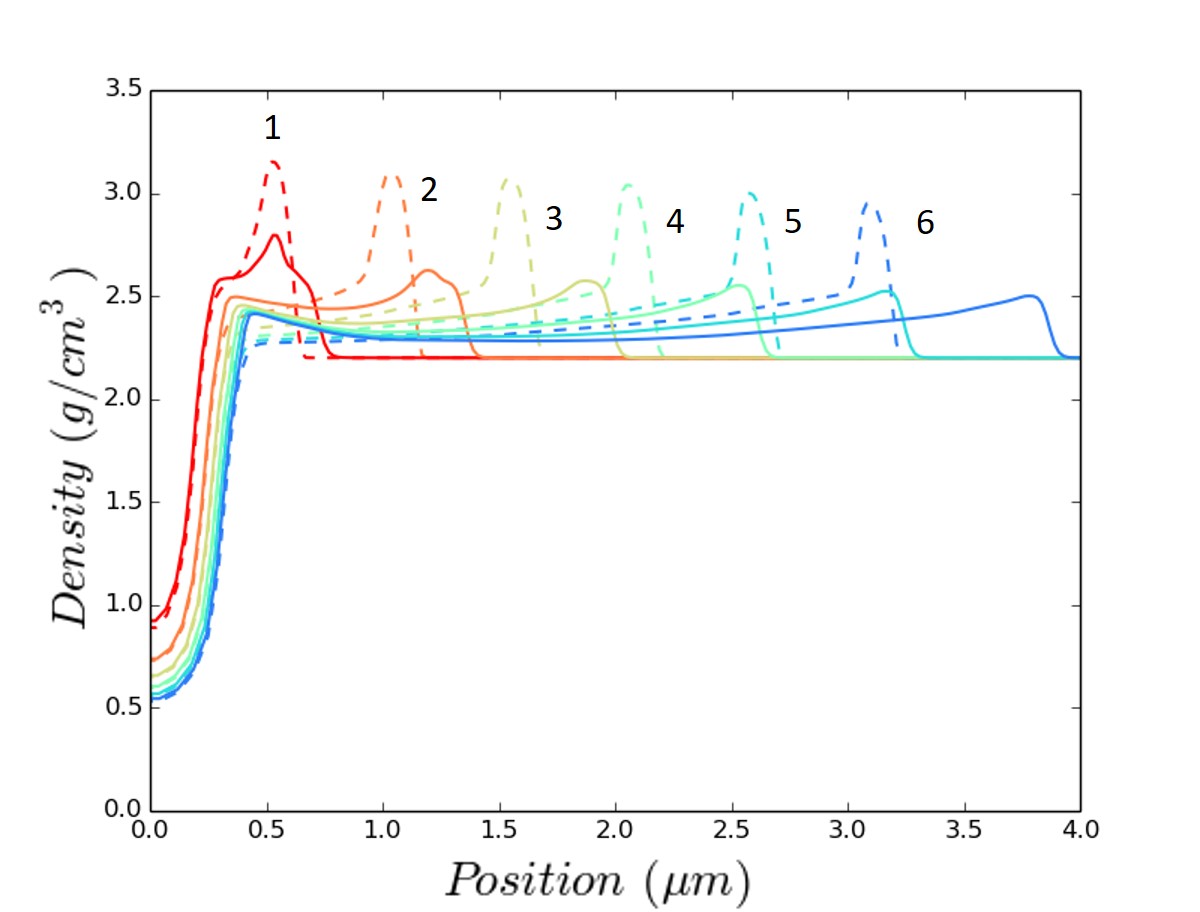}
  \caption{Spatial profiles of the density as predicted by the fluid model (dashed lines) and the EP model (lines), respectively, at
  100, 200, 300, 400, 500, 600 ps (curves 1-6, respectively).}
  \label{Figure_3}
\end{figure}

These differences between the fluid and EP behaviors have consequences on the density evolution as shown in Fig. \ref{Figure_3}, where density profiles
are presented for various times.
First, the formation of a cavity (from a density decreased by a factor two) is observed in both cases. The compression of the material by the shock wave induces then a strong relaxation (decompression) of the previously
compressed matter behind it. This leads to rarefaction waves which digs the material. 
The major difference appears in the surrounding matter where an overdense zone is formed in the case of the EP model.
With a strong shock front, it is possible to increase enough the internal energy of matter to liquefy it locally. 
This phenomenon can be observed from 100ps to 200 ps in the case of the EP model. 
The shock front always experiences the matter in the solid phase.
It induces an internal energy in excess of the energy of melting, inducing a liquid phase which is experienced by the back of the shock.
Since the speed of sound in the liquid phase is smaller than in the solid, the back of the shock is decelerated leading to the overdense region.

\begin{figure}[!h]
  \centering
  \includegraphics[width=7cm]{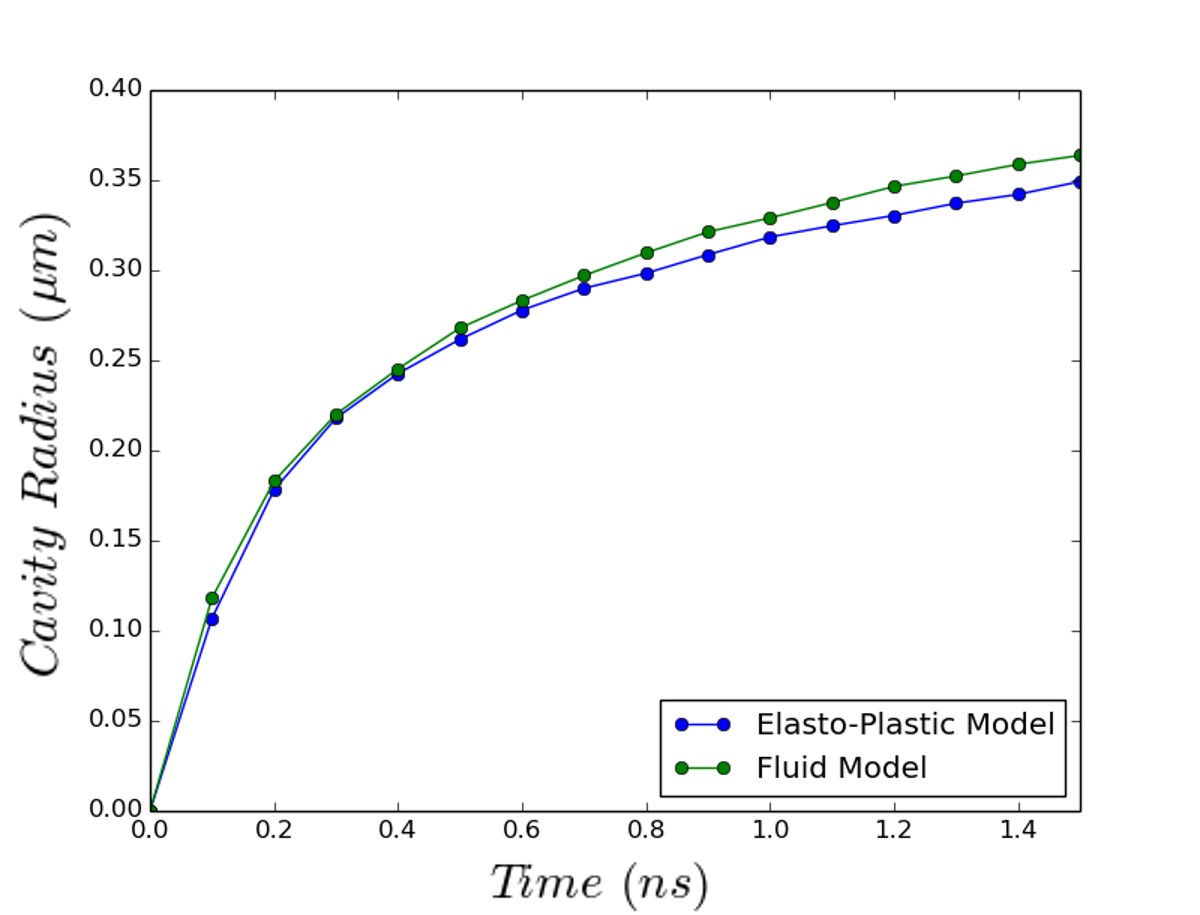}
  \caption{Evolution of the cavity radius as a function of time.}
  \label{Figure_4}
\end{figure}

The evolution as a function of time of the cavity radius is plotted in Fig. \ref{Figure_4}. 
This radius is defined as the distance from the center of the cavity, where the density is the lowest, 
to the position where the density is equal to half of the initial density. We have checked that a slightly different criterion leads to similar conclusions.
For both models the cavity radius increases with respect to time. Up to roughly 200 ps,
the cavity expansion is supported by the presence of the shock, leading to an almost constant speed of expansion for both models.
When the shock has traveled sufficiently far, the flow velocity decreases in the central region, explaining the observed slower expansion of the cavity.
In the case of the EP model, the cavity grows slightly slower than in the fluid case due to an additionnal force derived from the deviatoric part of the stress tensor opposed to the matter displacements in the solid phase.

To summarize, this 1D study has first shown that the shock divides into two parts due to the EP response. Secondly, the associated rarefaction wave creates a cavity which is surrounded by an overdense area. For the studied timescale,
the cavity radius increases monotically with respect to time without exhibiting any significant saturation whereas a finite size is expected based on experimental observations. An additional spatial dimension, 
by adding shear stress, is then expected to modify the cavity expansion dynamics as shown hereafter.
  
\subsection{2D simulation of cavity formation}
\label{2D}
Let us now consider the 2D case. The full cartesian mesh size is set to 18 $\mu$m$\times$18 $\mu$m with initial cell size set to 50 nm$\times$50 nm (simulations with smaller initial cells size were 
already performed to verify the numerical convergence). All parameters are similar to the previous section except the energy density 
which is set to 0.6 MJ/cm$^3$ (6 nJ) in the center of the 2D domain
to get a final cavity radius around 0.4 $\mu$m. Note that this absorbed 
energy density is different from the 1D case due to the additional spatial dimension.
With this energy deposition, the EOS predicts initial pressure and temperature around 0.3 TPa and $10^5$ K, respectively, in the energy deposition area.

\begin{figure}[!h]
  \centering
  \includegraphics[width=5cm]{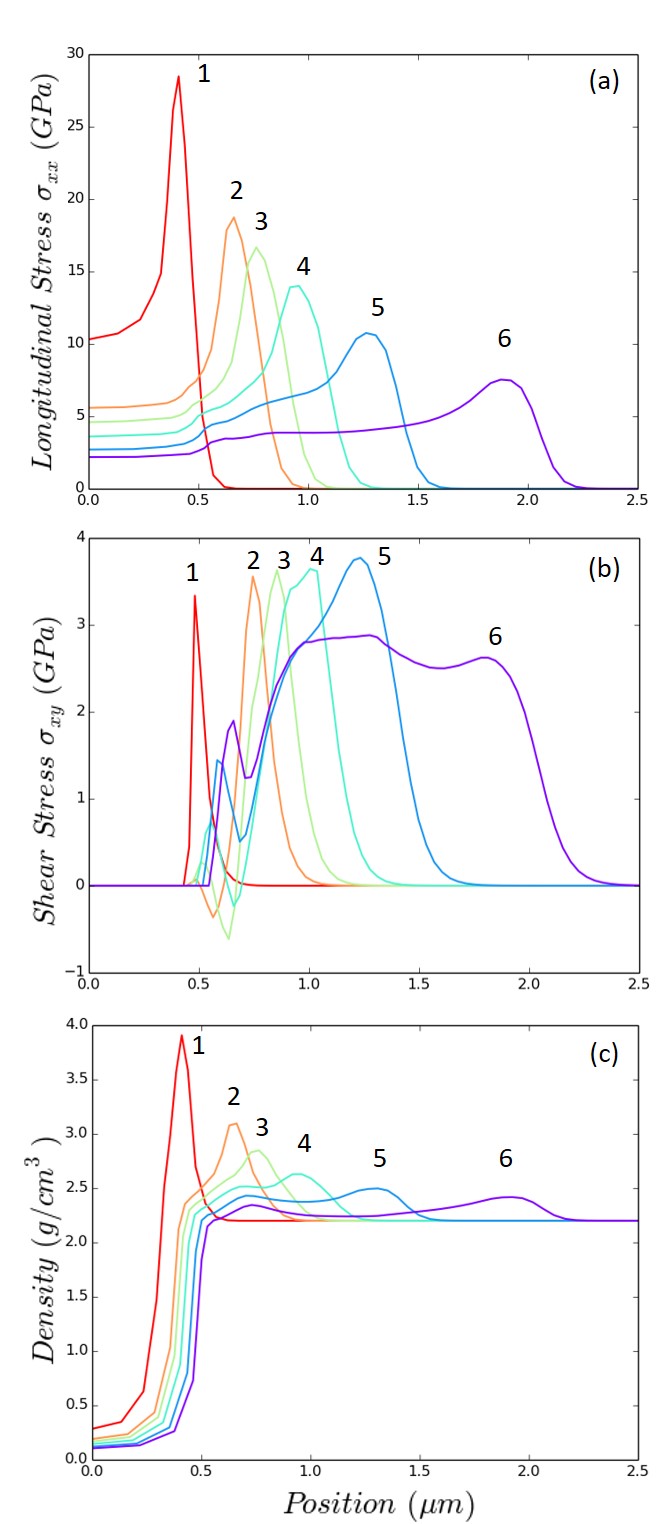}
  \caption{Evolution of (a) the longitudinal stress $\sigma_{xx}$,  
  (b) the shear stress $\sigma_{xy}$ and (c) the density profiles for times 50, 100, 120, 150, 200, 300 ps(curves 1-6, respectively)}
  \label{Figure_5}
\end{figure}
Figure \ref{Figure_5} presents the evolutions of (a) $\sigma_{xx}$, (b) $\sigma_{xy}$ and (c) the density profiles for various times. 
As in the 1D case, the shock first propagates with a straight front in the liquid phase (Fig. \ref{Figure_5}(a), roughly before 100 ps). After 100 ps (the temperature in the center of the
heated matter is around several 10$^4$K), the shock front experiences the solid state but induces a phase transition 
to the liquid state due to its large
amplitude. Since the sound velocity in the liquid is smaller, the back of the shock is decelerated leading to a change in the stress shape(as explained in the 1D case) with a split between the back and the front of the shock.
The elastic precursor is not really visible for this energy deposition but its signature can be seen around $\sigma_{xx}=Y_{HEL}=12$ GPa between 100 and 200 ps where the front shape slightly changes.
Figure \ref{Figure_5}(b) exhibits an accumulation of the shear stress $\sigma_{xy}$ during 200 ps around the cavity and following the front of the compression wave. A decreasing of the shear stress is visible behind the wave
(negative during 150 ps) due to the relaxation.
After 200 ps, the shear stress still increases around the cavity
but decreases while the shock transforms into an acoustic wave. Between this two regions, residual shear stresses (plateau) is created,  signature of the induced permanent deformations.
Figure \ref{Figure_5}(c) shows that a cavity is formed 
due to the strong relaxation behind the shock. As in the 1D case, an overdense region is induced in the vicinity of the cavity for the same reasons.
The final thickness (at 1.4 ns) of the overdense region, depending on the plastic deformation, is between 0.3 and 0.4 $\mu$m (the temperature at 1.4 ns in the center of the cavity is still high, around 10$^4$ K). 
Note this thickness may slightly change depending on the choice of the yield criterion 
(because the plasticity region change) \cite{Coulomb1776, Tresc1864, Taylor1934, Kermouche2008}. However, all the main trends presented above should be similar.
For the present purpose of laser structuration of materials, the von Mises yield criterion is well adapted.
Other criteria and models could be more suited for advanced studies, as the accurate description of the hardening in the densification process of SiO$_2$ for instance \cite{Kermouche2008}.

\begin{figure}[!h]
  \centering
  \includegraphics[width=8cm]{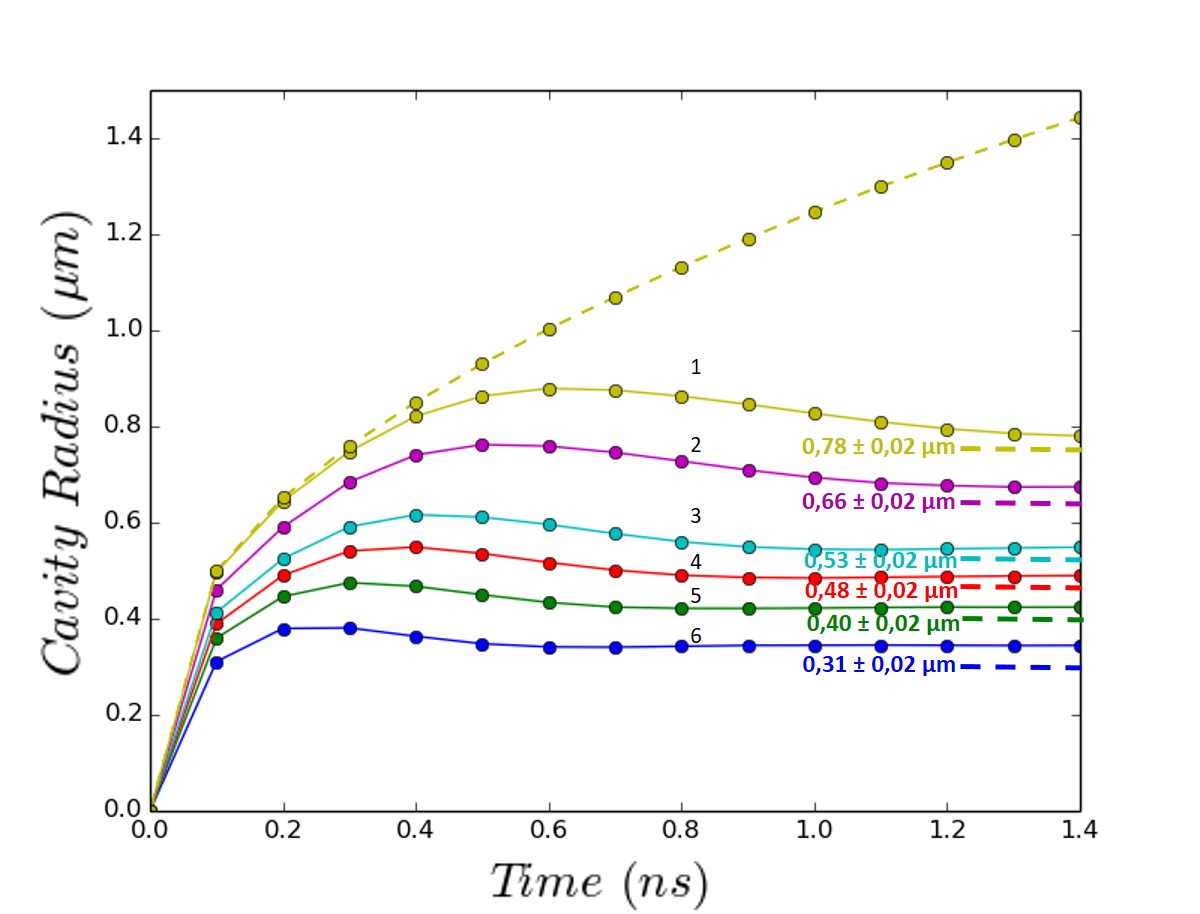}
  \caption{Evolution of the cavity radius as a function of time for different energy densities: 2, 1.5, 1, 0.8, 0.6, 0.4 MJ/cm$^3$ (solid curves 1-6, respectively) and comparison with the 
  fluid model for 2 MJ/cm$^3$ (yellow dashed curve). Predictions (see text) of final cavity radius 
  (horizontal dashed lines) are provided in the case of the EP model.}
  \label{Figure_6}
\end{figure}

Evolutions of the cavity radius as a function of time for various initial energy densities, predicted by the EP model, are plotted in Fig. \ref{Figure_6}. 
An evolution for 
the largest energy density is also simulated with the fluid approach for the sake of comparison. Within the EP model, the evolution of the cavity radius consists mainly of three
steps: an expansion, a contraction (shrinking), and a stabilization.
The first step is due to the formation of a strong shock as in the 1D case. During the expansion, shear stress is accumulated (Fig. \ref{Figure_5}(b)), inducing an additional force opposing to the matter displacement. 
Ultimately, the expansion stops when the shock transforms into a pure elastic wave. The cavity radius then reaches a maximum value. It is worth noting that this behavior takes place on a shorter timescale than the 1D case due to the 
influence of the shear stress which is more significant in the 2D simulations. Then, corresponding to the above-mentioned second step, the cavity radius decreases due to the relaxation of the elastic deformations closing partially the cavity. 
Finally, the cavity stabilizes (third step) due to the induced permanent deformations.That is this final radius which has to be compared to
experimental observations in post-mortem analysis.

The evolution of the previous trends with respect to the absorbed energy density is now discussed. During the first step, the larger the absorbed energy density, the stronger the shock, and the faster
the expansion of the cavity. That leads to a maximum cavity radius which increases and is reached for longer times. For the second step, the larger the energy absorption, the longer the elastic relaxation time 
to reach the stabilization of the cavity. Indeed, elastic deformations are more important.
The final radius increases as a function of the absorbed energy density; its behavior is discussed in more details hereafter.

The final radius can also be estimated from the reached maximum value by subtracting the total elastic deformations. Within this procedure, we take into account the possible phase changes which remove
the solid deformations. As shown by 
Fig. \ref{Figure_6} (horizontal dashed lines), results of these predictions are in a good agreement with the final radius as predicted by the full EP modeling. This result supports the
analysis performed in the previous paragraph.

Figure \ref{Figure_6} also presents the evolution of the cavity radius in the case of the fluid model for an absorbed energy density of 2MJ/cm$^3$. The radius increases monotonically
without exhibiting any decrease, neither stabilization as observed with the EP model, demonstrating the importance of the latter. However, a final cavity radius may be determined by comparing
the front shock pressure $P_{shock}$ to the Hugoniot elastic limit, \textit{i.e.} it is obtained when $P_{shock}=Y_{HEL}$. In the present case, that takes place at 480 ps, leading to an estimation around
0.9 $\mu$m. This value corresponds approximatively to the maximal radius given by the EP model. However, the fluid model is not able to exhibit the elastic relaxation (second step) as the EP model which predicts a 
smaller final cavity radius of 0.8 $\mu$m or so. 

\begin{figure}[!h]
  \centering
  \includegraphics[width=8cm]{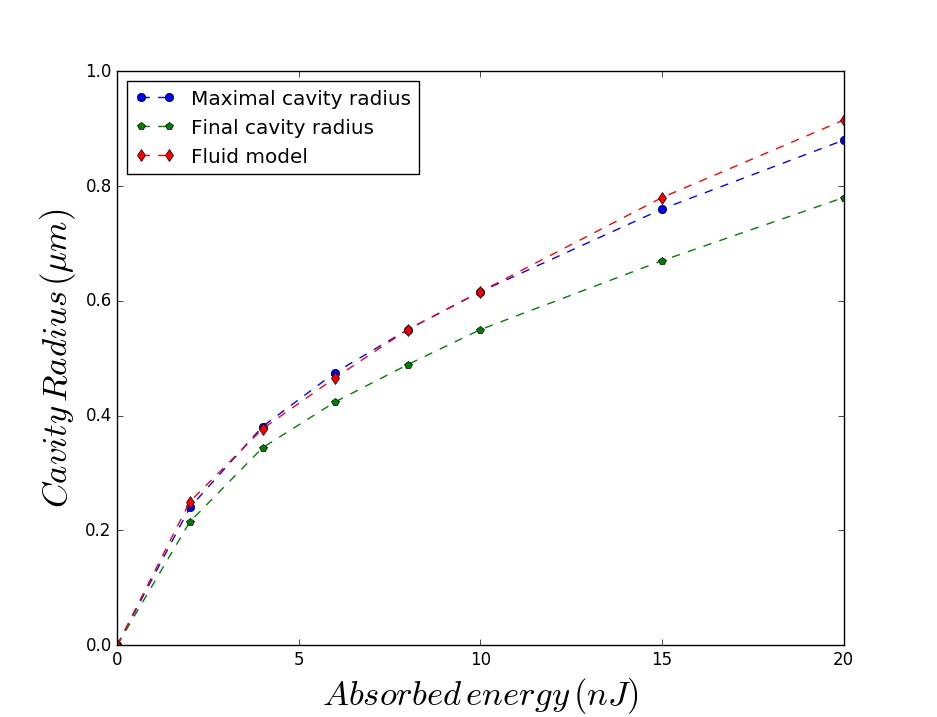}
  \caption{Evolution of the maximum and final cavity radius as a function of the absorbed energy density.}
  \label{Figure_7}
\end{figure}

Figure \ref{Figure_7} presents the evolution of the maximal and the final cavity radius as a function of the absorbed energy within the EP and fluid models.
With an absorbed energy in a range from 2 to 20 nJ (0.2 to 2 MJ/cm$^3$), the final cavity radius is about 0.2 to 0.8 $\mu$m within the EP model. It increases with the absorbed energy as a square root (checked by a fit),
\textit{i.e.} $r_{cavity}\propto \sqrt{E_{abs}}$,
due to the 2D geometry (for a 3D geometry, $r_{cavity}\propto \sqrt[3]{E_{abs}}$). The difference between the maximal cavity radius and the final cavity radius increases with the absorbed energy.
This is due to the increase of the region size where the matter has been compressed. The elastic relaxation then leads to a larger difference between transient and final radius. Within the fluid model, 
as mentionned above, the final cavity radius corresponds to the maximal radius reached by the EP model, due to the absence of elastic behavior.
Thus the discrepancies exhibited by the fluid model in both the final value and the temporal behavior of the cavity radius clearly shows the importance of the EP behavior.
For a more realistic energy deposition, especially for a larger absorption volume possibly corresponding to a higher energy of the laser beam, a change in the temporal evolution is expected, but not necessarily in the final cavity size.
Indeed, the energy density should be lower due to the increase of the absorption volume leading to a lower pressure and a lower temperature. Thus, the shock wave should be weaker and should induce a smaller increase
of the cavity radius accompanied by a weaker plastic deformation. This should lead to a lower final cavity size. However, the larger absorption zone should lead to a
larger final cavity size. Thus, both influences are in competition and should lead to comparable final cavity sizes.  
The improvement of the EP approach is also its ability to predict the potential crack formation as shown hereafter.

\subsection{Predictions of potential cracks formation}
This modeling permits to determine different critical areas where potential cracks (fractures) may appear. 
In the case of ideal and homogeneous materials without any defects, the resistance corresponds to the pressure capable
of separating atoms \cite{Griffith1921}, leading to very large value of the material resistance. In the realistic case, the resistance (ability to support stress) is significantly decreased by the presence of defects which initiate
the formation of crack. Effective macroscopic values of the intrinsic mechanical limits  of materials can then be defined. They consist of the resistance 
in compression $L_c$, traction $L_t$, and shear \cite{Spenle2003} $L_t/2$. $L_t$ is set to a characteristic value \cite{Pedone2015} (microscopic scale) of 8 GPa and $L_c$ is set to 10 GPa (resistance in compression 
is generally larger than resistance in traction 
for brittle materials \cite{Fanchon2001}).

The present EP simulations allows us to perform
the laser induced stress in the material. From it, principal stresses in compression-traction and maximum shear stress can be defined.
If the laser induced principal stress or maximum shear stress exceeds the associated previous limits, then the material may break, \textit{i.e.} 
cracks may appear. Details are provided in Appendix B.

\begin{figure}[!h]
  \centering
  \includegraphics[width=8cm]{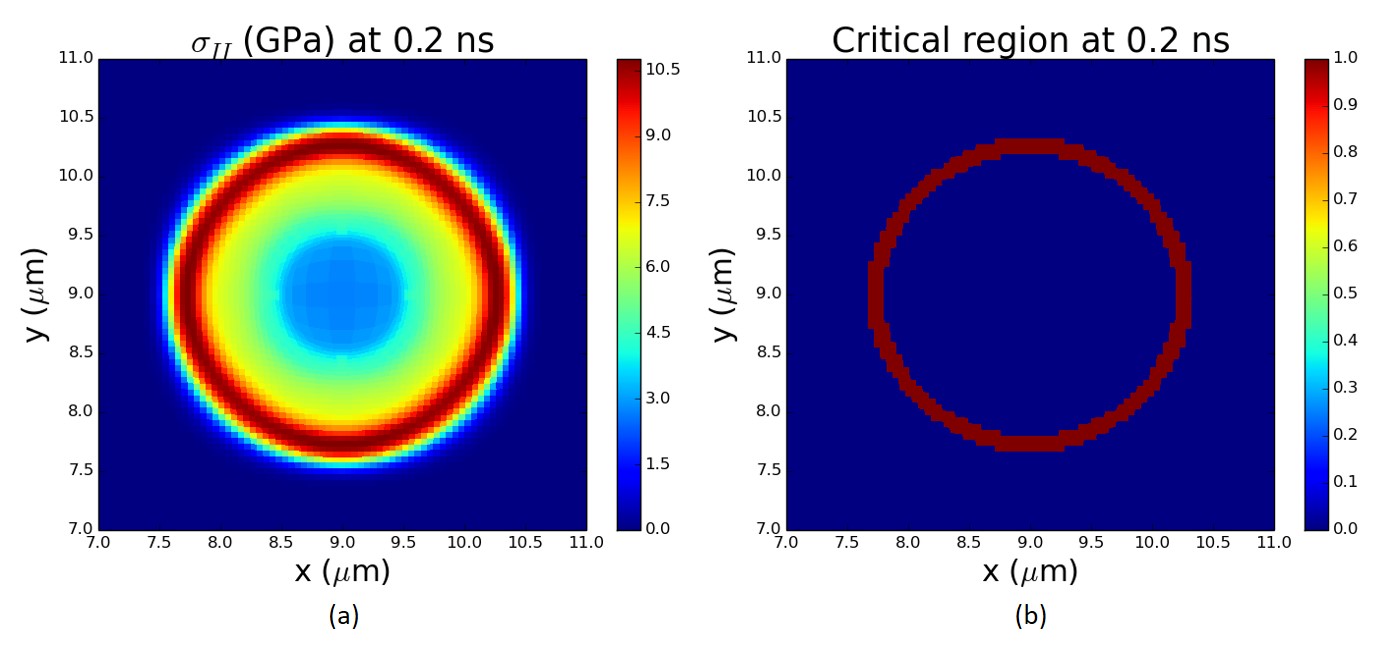}
  \caption{(a) 2D map of principal stresses in compression $\sigma_{II}$. (b) Critical zones in compression around the cavity.}
  \label{Figure_8}
\end{figure}

Within the conditions of the previous section, an evaluation of this critical area has been performed. In the present case of isotropic energy deposition and material
mechanical properties, the symmetry of the system leads to conclude that stress in compression is the main component. Figure \ref{Figure_8}(a) shows
its spatial distribution at 200 ps for an energy deposition of 0.6 MJ/cm$^3$.
At this time, the stress is maximal, \textit{i.e.} the crack formation is the most probable,
and located around the cavity. As expected, the stress is distributed isotropically with a maximum value for a radius slightly larger than 1 $\mu$m which is roughly twice the cavity radius (0.4 $\mu$m). This
induced stress leads to a critical area as shown in Fig. \ref{Figure_8}(b). It is distributed similarly as the previously discussed maximum stress, exhibiting a thin ring shape which the probability of meeting and activating of
defects (inducing a crack) depends on its size and the density of defects.
The density of activated defects can be defined as a function of their spatial distribution (Weibull modulus), their size and the stress applied to the material. The probability of crack initiation can then be 
evaluated through the Weibull distribution \cite{Rinne2008}. Furthermore a model of mechanical fracturation \cite{Reyne2015} could be investigated in order to charaterize the fracture length as a function of the cavity radius. 
However, such a study is out of the scope of the present work.

\begin{figure}[!h]
  \centering
  \includegraphics[width=8cm]{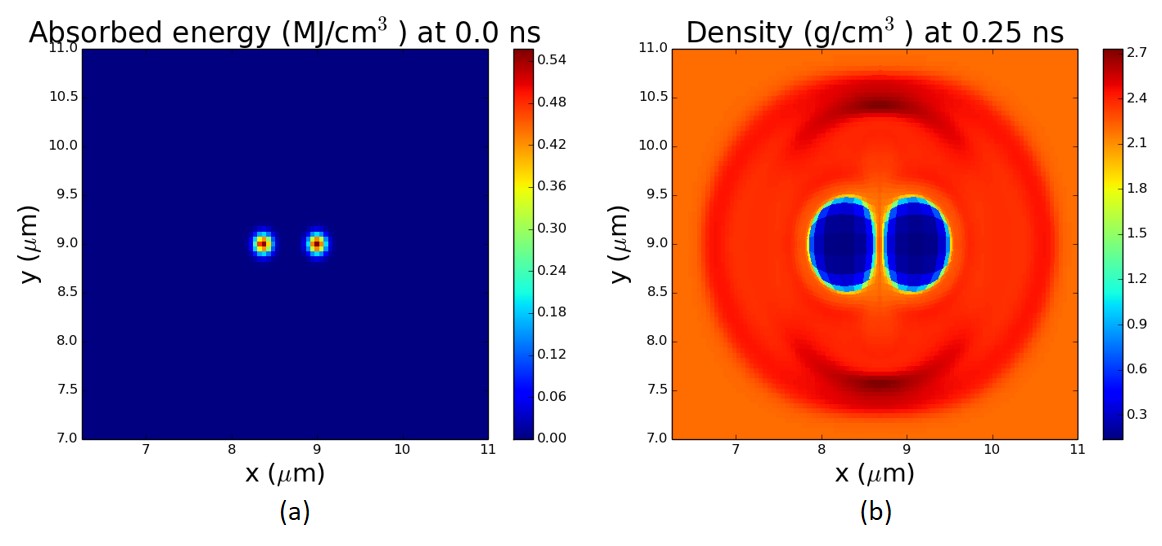}
  \caption{2D map of (a) the initial absorbed energy density and (b) the density profile with an inhomogeneous energy deposition at 250 ps.}
  \label{Figure_9}
\end{figure}
In order to exhibit the influence of the shear stress, an inhomogeneous energy density deposition is now considered. 
The total energy is set to 12 nJ ($1.2 MJ/cm^3$ in average) which is twice the previous energy,
and distributed within two main spots separated by a few tenths of micrometers, (see Fig.\ref{Figure_9}(a)). Figure \ref{Figure_9}(b) shows the induced 2D density profile at 250 ps.
Two cavities, separated by a slightly denser region, corresponding to the main energy deposition spots are created.
As in the previous homogeneous case, a shock wave launching is associated to each cavity formation. In the horizontal direction outwards the cavities, the shock waves mainly
propagate as in the homogeneous case. However, for the inward direction, the shock waves collide, leading to a daughter wave in the perpendicular direction. That results in
two symmetric overdense regions in the vertical direction as depicted by Fig. \ref{Figure_9}(b).

\begin{figure}[!h]
  \centering
  \includegraphics[width=8cm]{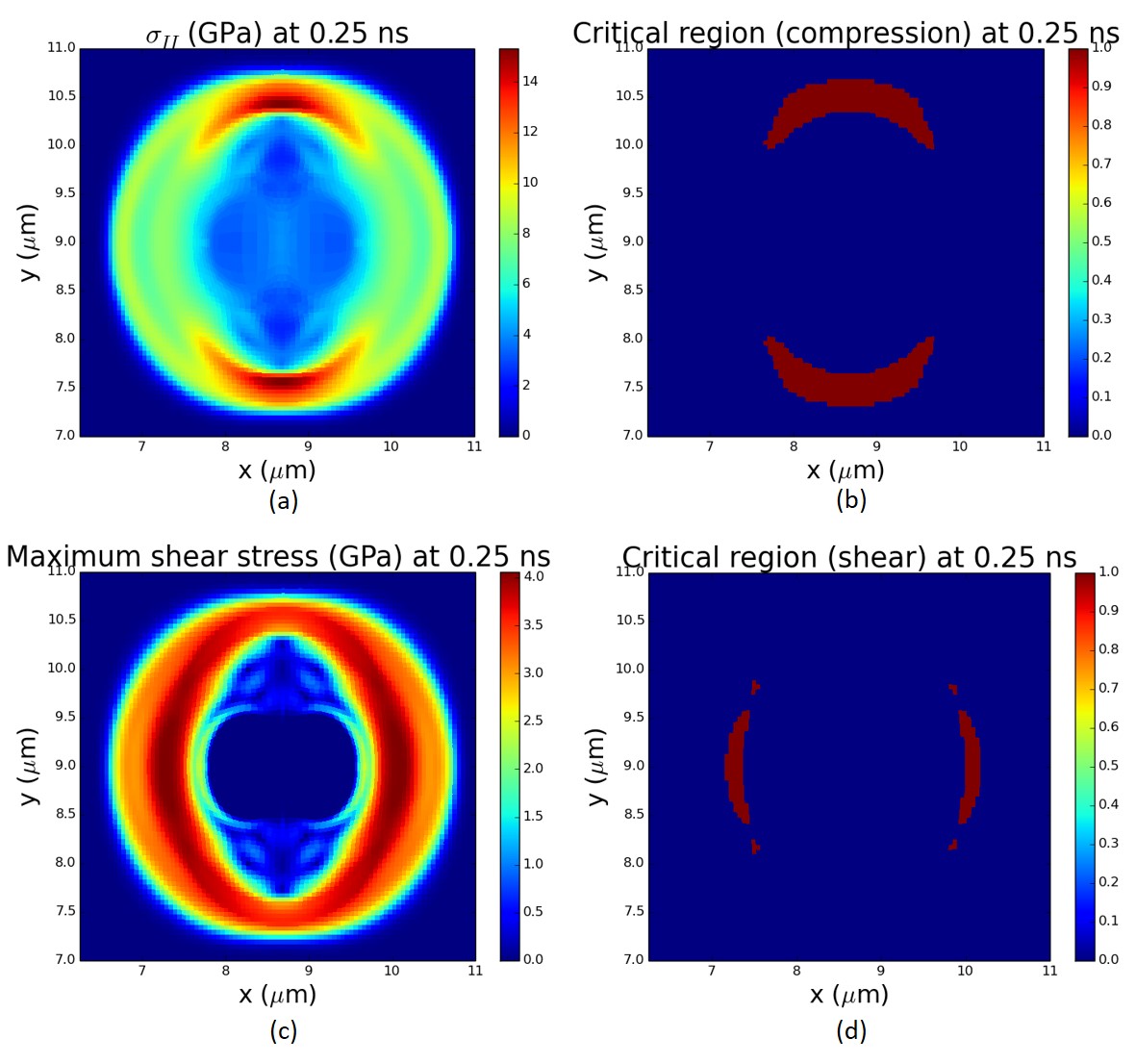}
  \caption{2D spatial profiles at 250 ps of (a) the principal stress in compression with (b) its associated critical area; (c) the maximum shear stress with (d) its associated critical area.}
  \label{Figure_10}
\end{figure}

Figure \ref{Figure_10} shows the principal stress in compression
(Fig. \ref{Figure_10}(a)) and the maximum shear stress (Fig. \ref{Figure_10}(c)) together with their associated critical areas (Figs. \ref{Figure_10}(b) and (d), respectively) at 250 ps.
As discussed for the density map, the daughter shock wave creates the maximum compressive stress in the vertical direction, resulting in both the upper and lower regions of maximal stress.
That leads to critical areas exhibiting a similar appearance as depicted by Fig. \ref{Figure_10}(b). The overall clearly shows the anisotropic distribution of stress. The latter then gives
rise to an elongation in the vertical direction leading to the apparition of shear stress on the left and right sides of the double-cavity structure, as shown by Fig. \ref{Figure_10}(c). The critical sheared region also
exhibits the same anisotropy with a main axis oriented vertically. The potential formation of cracks is thus expected to take place in a preferential direction due to the initial anisotropic energy
deposition. We also expect a crack propagation in a preferential direction. These numerical results are in agreement with a recent experimental study \cite{Zimmermann2016} devoted to a similar physical system. The observations have shown
the same trends, \textit{i.e.} an elongation of the created cavities and preferential cracks directions.

\section{CONCLUSION}
\label{conclusion}
A theoretical and numerical modeling has been presented to analyze and understand the formation of structural modifications in transparent materials induced by a
femtosecond laser pulse tightly focused in the bulk. The standard fluid description has been augmented by the solid response through the elasto-plastic behavior.
This approach has been implemented in the 2D hydrodynamic CHIC code. Numerical simulations have been performed for fused silica irradiated within such
conditions that a cavity may form. It has been shown that the creation of a shock wave and its subsequent propagation lead to a cavity formation and surrounding overdense region.
Due to the elasto-plastic behavior, the shock exhibits a two-wave structure: an elastic precursor inducing reversible deformations and a plastic wave leading to permanent deformations. 
The induced shear stress contributes significantly to the dynamics of the cavity formation, ultimately leading to stop its expansion. The maximum cavity size is reached when the shock transforms into a pure elastic wave. 
The relaxation of the elastic deformations then closes partially the cavity. The permanent deformations are responsible for the final cavity stabilization and the final overdense area. 
By taking the same order of magnitude of the absorbed energy deposition as experimentally obtained \cite{Gamaly2006},
the present approach, as it is, permits to account for the morphology of the observed cavities.

Such behaviors cannot be obtained by the fluid approach. In particular, the latter cannot lead to any cavity stabilization which size could be compared to experimental
observations. To overcome this flaw, by comparing the hydrodynamic pressure to the Hugoniot elastic limit, an ad hoc criterion renders it possible to give the order of magnitude
of the final cavity radius, however without being able to describe the surrounding matter. It turns out that
a reliable comparison to experimental data is only possible by accounting for the elasto-plastic influence.

By evaluating the induced mechanical stress, the present approach is also able to predict the regions where potential fractures may appear during the interaction.
With an isotropic energy deposition, potential fractures are mainly induced by a compressive stress. Thanks to the present 2D approach,
anisotropic conditions can be considered. In addition to the compression, an anisotropic energy deposition is shown to possibly create fractures from shear stresses.

Since the present approach is general, simulations for other materials could also be performed by adapting the material properties as the equation of state, the mechanical constants, etc.
This thermo-elasto-plastic tool could thus be used to design laser-induced nano-structures in material within conditions preventing from formation of fractures
(which are detrimental to the structure functionalities).

Finally, the simulation of the laser propagation, including the electrons dynamics, will be performed to obtain a more realistic energy deposition.
A 3D modelling should also be developed to perform more quantitative predictions
which may account accurately for experimental observations.
Since the main physical mechanisms are expected to be the same as those exhibited in the present work, the present modeling and conclusions may be used as a baseline for future developments
in the field of the laser-matter interaction.

\section{ACKNOWLEDGMENTS}
Vladimir Tikhonchuk is acknowledged for fruitful discussions. The CEA and the R\'egion Aquitaine (MOTIF project) are acknowledged for supporting this work. Jocelain Trela is also akcnowledged
for the help brought on plots of 2D maps. 

\begin{appendices}

\section{\textbf{Thermal Softening}}
\label{annexesA}
The solid-liquid phase change is modelized by a thermal softening model. A polynomial $g$($\varepsilon$) weights the shear modulus and the yield strength. It is equal to unity and zero
for the solid and liquid phases, respectively, depending on the internal energy $\varepsilon$. A third order polynomial is chosen to ensure a smooth transition between the two states. More precisely,
it is defined as follows. If $\varepsilon<\varepsilon_{solid}$, then $g(\varepsilon)=1$.
If $\varepsilon>\varepsilon_{solid}$ and $\varepsilon<\varepsilon_{melting}$, then:
\begin{equation}
 g(\varepsilon)=1-3(\frac{\varepsilon-\varepsilon_{solid}}{\varepsilon_{melting}-\varepsilon_{solid}})^2+2(\frac{\varepsilon-\varepsilon_{solid}}{\varepsilon_{melting}-\varepsilon_{solid}})^3
\end{equation}
And if $\varepsilon>\varepsilon_{melting}$, then $g(\varepsilon)=0$.
$\varepsilon_{solid}$ is the internal energy of the solid at 300 K ($\varepsilon_{solid}=0.3$ kJ/g) and $\varepsilon_{melting}$ is the internal energy of melting ($\varepsilon_{melting}=2.15$ kJ/g, corresponding to a softening temperature
around 2000 K \cite{Meshcheryakov2013}).

\section{\textbf{Stresses analysis}}
\label{annexesB}
\begin{figure}[!h]
  \centering
  \includegraphics[width=6cm]{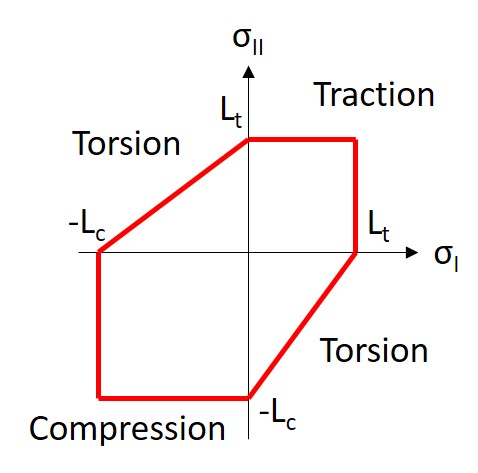}
  \caption{Shematization of the Mohr's criterion}
  \label{Figure_11}
\end{figure}

In order to perform the analysis of plane stresses, the knowledge of principal stresses in every points of the material is required. The matrix elements of $\bar{\bar\sigma}$ in the 2D cartesian directions of coordinates system are 
defined as $\sigma_x$, $\sigma_y$ and $\tau_{xy}$, the longitudinal stresses and the shear stress, respectively. They are the projection on the two axis of applied stresses. Knowing these values in these particular directions, 
other projections of stresses 
defined as $\sigma_{x1}$, $\sigma_{y1}$ and $\tau_{x1y1}$ can be calculated in any directions making an angle $\theta$ with the abscissa axis \cite{Fanchon2001}.
\begin{equation}
 \sigma_{x1} = \frac{\sigma_x+\sigma_y}{2}+\frac{\sigma_x-\sigma_y}{2}\cos{2\theta}+\tau_{xy}\sin{2\theta}
\end{equation}
\begin{equation}
 \tau_{x1y1} = -\frac{\sigma_x-\sigma_y}{2}\sin{2\theta}+\tau_{xy}\cos{2\theta}
\end{equation}
\begin{equation}
 \sigma_{y1} = \frac{\sigma_x+\sigma_y}{2}-\frac{\sigma_x-\sigma_y}{2}\cos{2\theta}-\tau_{xy}\sin{2\theta}
\end{equation}
A particular angle $\theta_p$ exists, where $\sigma_{x1}$ is maximized and simultaneously $\sigma_{y1}$ is minimized, and $\tau_{x1y1}$ is canceled. This is the 
principal coordinate system which amounts to diagonalize the stress tensor. In this case, the matter is in a traction/compression state. It is thus interesting to use this
frame in order to compare principal stresses to traction/compression limits. $\sigma_I$ is defined as the maximum (minimum) stress and 
$\sigma_{II}$ as the minimum (maximum) stress in traction (compression). By convention $\sigma_I>\sigma_{II}$.
\begin{equation}
 \sigma_I =\sigma_{m}+R 
\end{equation}
\begin{equation}
 \sigma_{II} = \sigma_{m}-R
\end{equation}
with
\begin{equation}
 \sigma_{m} = \frac{\sigma_x+\sigma_y}{2}
\end{equation}
and
\begin{equation}
 R = \sqrt{(\frac{\sigma_x-\sigma_y}{2})^2+\tau_{xy}^2}
\end{equation}
The angle $\theta_p$ can be performed through the relation:
\begin{equation}
 \tan{2\theta_p} = \frac{2\tau_{xy}}{\sigma_x-\sigma_y}
\end{equation}
Similarly, a frame rotation with an angle $\theta_s$ can be defined, at $\pm45^\circ$ relative to the principal coordinates system (Mohr's circle \cite{Parry2004}), where the shear stress 
is maximized whereas normal stresses are equal to the mean stress $\sigma_{m}$. This frame is useful to know sites where the 
maximum shear stress may create dislocations or cleavages. The relations read:
\begin{equation}
 \tau_{max} = R
\end{equation}
and
\begin{equation}
 \theta_s = \theta_p \pm 45^\circ
\end{equation}
Finally, to determine critical sites, where stresses exceed intrinsic mechanical limits of material, Mohr's criterion is used \cite{Fanchon2001}. One can define a compression limit $L_c$ and a 
traction limit $L_t$, especially in breakable materials where they are differents.
Thus, thanks to the principal basis, $\sigma_I$ and $\sigma_{II}$ can be compared to these limits deducing critical points. If $\sigma_I$ 
and $\sigma_{II}$ have the same sign, the largest one must not exceed $L_c$ or $L_t$ depending on the stress state. In the case of a traction state, stresses are positives, $\sigma_I$ 
represents the maximum stress and $\sigma_{II}$ the minimum stress in traction. Within this condition, a point is critical if 
\begin{equation}
 \frac{\sigma_I}{L_t}\geq1
\end{equation}
In the opposite case of a compression state, a point is critical if 
\begin{equation}
 \frac{|\sigma_{II}|}{L_c}\geq1
\end{equation}
If $\sigma_I$ and $\sigma_{II}$ have opposite signs, that corresponds to torsion. A point is critical if:
\begin{equation}
 \frac{|\sigma_I-\sigma_{II}|}{L_t}\geq1
\end{equation}
or
\begin{equation}
 \frac{|\sigma_I-\sigma_{II}|}{L_c}\geq1
\end{equation}
which is illustrated in Fig.\ref{Figure_11}.

\end{appendices}

%

\end{document}